\documentstyle[pra,aps,preprint,floats,psfig]{revtex}
\input epsf
\def\eqr#1{{Eq.~(\ref{#1})}}

\def\grtsim{\,\,\rlap{\raise 3pt\hbox{$>$}}{\lower 3pt\hbox{$\sim$}}\,\,}
\def\lsim{\,\,\rlap{\raise 3pt\hbox{$<$}}{\lower 3pt\hbox{$\sim$}}\,\,}

\begin{document}
\preprint{\vbox{\baselineskip=12pt
\rightline{SLAC-PUB-8654}
\vskip0.2truecm   
\rightline{hep-ph/0010103}}}

\title{Experimental Probes of the Randall-Sundrum Infinite Extra
Dimension}
%\vspace{0.5in}
\author{Daniel J.\ H.\ Chung\thanks{Electronic mail:
djchung@umich.edu} and Lisa Everett\thanks{Electronic mail:
leverett@feynman.physics.lsa.umich.edu}}
\address{Randall Physics Laboratory,
     University of Michigan, Ann Arbor, Michigan \ 48109-1120}

\author{Hooman Davoudiasl\thanks{Electronic mail:
hooman@slac.stanford.edu}}
\address{Stanford Linear Accelerator Center, Stanford University,
Stanford, CA 94309}

%\vspace{0.5in}
\maketitle

\begin{abstract}

The phenomenological possibilities of the Randall-Sundrum non-compact
extra dimension scenario with the AdS horizon increased to
approximately a millimeter length, corresponding to an effective brane
tension of $(\rm TeV)^4$, are investigated. The corrections to the
Newtonian potential are found to be the only observationally
accessible probe of this scenario, as previously suggested in the
literature.  In particular, the presence of the continuum of KK modes
does not lead to any observable collider signatures.  The extent to
which experimental tests of Newtonian gravity can distinguish this
scenario from the scenario of Arkani-Hamed, Dimopoulos, and Dvali with
one and two millimeter size extra dimensions is explicitly
demonstrated.

\end{abstract}

\pagebreak
%%%%%%%%%%%%%%%%%%%%%%%%%%%%%%%%%%%%
%\section{Introduction}
%%%%%%%%%%%%%%%%%%%%%%%%%%%%%%%%%%%
Much of the recent excitement and theoretical speculation regarding
large/warped compact extra dimensions found for example in
Refs.~\cite{ADD,antoniadis,ddg,RS1} 
has been fueled by the
amelioration/novel rephrasing of the various hierarchy problems. In
addition, the models
of Refs.~\cite{ADD,RS1} have raised some hopes that upcoming
experiments may reveal signatures of the existence of higher
dimensions \cite{constraints,schmaltz,dhr}.
%and in earlier work \cite{akama,rubakov} 
On the other hand, the novelty of
having phenomenologically consistent four-dimensional gravity in the
presence of non-compact higher dimensions has been the main motivation
for the investigations of ``warped bulk'' models based on the idea of
Randall and Sundrum \cite{RSII,hybridgravity} without much motivation
from upcoming experimental signature possibilities.  In this paper, we
investigate the possible experimental signatures of the model of
Randall and Sundrum of Ref.\cite{RSII} in the case that the warping
scale is sufficiently lowered.  Since most of the qualitative aspects
of this scenario have already been discussed in Ref.~\cite{ADDK} and
Ref.~\cite{giddings}, this paper will focus on the details of the
experimental signatures.

Unlike the non-warped scenarios and some of the variations of the
warped model for which the gravity is quasilocalized (see also
\cite{hybridgravity,hybridcollider,largedistancepheno}), the original
warped bulk model of Randall and Sundrum \cite{RSII} (henceforth
referred to as RS) did not seem to have any low energy
phenomenological implications (perhaps with the exception of black
hole physics\cite{blackholes,garrigatanaka,generalstudy,giddings}), mainly
because the cosmological constant contribution coming from the
constant energy density on the Planck brane had always been identified
with the string scale for naive naturalness reasons.  (We will refer
to this constant energy density as the RS brane tension.)  However, as
explicitly noted for example by Kraus\cite{kraus}, if one attempts to
identify the RS brane tension with that of a collection of D3 branes,
one finds a discrepancy: the D3 brane tension is 2/3 of the brane
tension needed for the RS brane tension.

More recently, a different class of SUGRA solutions \cite{justsugra}
(albeit singular) have been discovered which share the warped bulk
spacetime of the RS scenario (although the noncompact limit has not
been explored).  Even more recently, there has also been progress in
embedding the SUGRA containing the RS solution within a particular
compactification of Type IIB string theory\cite{connectiontostring}.
One emerging picture is that the RS brane does not correspond to any
particular D-brane but is an ``effective geometry'' arising from a
combination of a stack of negative tension branes stuck at an orbifold
fixed point. Importantly, the positive quantity that was previously
naively identified with the brane tension is not really the brane
tension, but is actually a term arising from the combination of the
positive curvature from the orbifold fixed point singularity and the
negative tension of the branes confined there.  In this picture, the
RS brane tension is of the order of
\begin{equation}
V_{brane} \sim M_{pl}^2 k^2, 
\label{eq:branetension} 
\end{equation} 
where $1/k$ is the AdS horizon of the RS brane embedding spacetime (the
warping scale). It is given by \cite{kraus,connectiontostring}
\begin{equation} 
\frac{1}{k}=\frac{(4 \pi g N)^{1/4}}{M_{st}}, 
\end{equation}
where $M_{st}$ is the string scale, $g$ is the string coupling, and $2 N
\gg 1$ is the number of D3-branes stacked to form the RS brane. Hence, if
$N$ is taken large enough, the warping scale can be lowered to $k^{-1}
\sim 0.1$ mm such that the RS brane tension will be at $O$(1 TeV),
which may be a useful scale for model building (such as for
particle/sparticle mass splitting).  Additionally, given that the
cosmological constant scale is $O(10/\mbox{mm})^4$, it may not be
unreasonable to expect the $10/\mbox{mm}$ scale to enter the effective
theory. In any case, the RS model with small warping will predict
signatures for the upcoming experiments testing submillimeter behavior of
Newton's law. We will refer to this model as the $\overline{\rm RS}$
model.\footnote{ Although we restrict our investigations in this paper to
the $\overline{\rm RS}$ model, our conclusions should easily extend to its
generalizations such as that of Ref.~\cite{higherd}.}

As is well known, the leading order multiplicative correction to
Newton's law has a functional behavior of $1+1/(k^2 r^2)$ for the RS
model at large distances.  Hence, one would naively expect this scenario
to behave just like that of Ref.~\cite{ADD} (ADD model) with two extra
dimensions (n=2), which would imply that there would be collider
signatures for the $\overline{\rm RS}$ model.
On the other hand, as pointed out by Ref.~\cite{ADDK}, this
$1+1/(k^2 r^2)$ correction becomes $1/(k r)$ at distances much shorter
than the AdS horizon length of $1/k$.  In other words, the
$\overline{\rm RS}$ model for $\sqrt{s} \gg k$ is effectively ADD (n=1),
since at large $\sqrt{s}$ the curvature can be ignored. Hence, at
collider distances ($1/\mbox{TeV}$), for $k\sim (0.1\,\mbox{mm})^{-1}$ the
corrections from the continuum of KK states do not enhance the
gravitational processes sufficiently to cause them to be observable.

However, it is not clear that there will be absolutely no effects in
this case because the brane tension given by \eqr{eq:branetension} is
at the collider scale.  It can be argued that since the RS brane is
confined to an orbifold fixed point, the brane should not behave as a
thick or ``fat'' brane\footnote{For aspects of thick brane physics,
see for example 
\cite{thickbranegravity,schmaltz,thickbranecollider,aaron}.} even
when the collider energy is above the scale of the brane tension.  On
the other hand, it is not obvious whether the stacking of such a large
number of D-branes is consistent with the orbifold fixed point
idealization.  If the brane can fluctuate at the collider scale, the
brane tension may serve as a cutoff for the standard model theory
confined to the brane and may help solve the hierarchy problem (in
analogy with the lowering of the fundamental scale in the ADD
scenarios).

Even independently of the string theory picture (note that the string
theory realization of the RS model is not known to be unique) and
assuming somehow that a field theory can be well defined in the presence
of orbifold fixed point singularities, the RS brane tension may not always
reflect the scale of the tension of the object sitting at the orbifold
fixed point.  For example, suppose an extended object of tension $T_b$ is
sitting at the orbifold fixed point but is not confined to the fixed point.
Given that gravity itself is a derivative expansion of a more fundamental
theory and that the curvature at the orbifold fixed point is singular, one
would expect there to be large corrections to Einstein's equation
arising from higher derivative curvature terms.  Hence, one would find
that the effective RS brane tension $T_{RS}$ is a sum of an infinite
number of higher curvature terms and $T_b$: e.g. schematically,
\begin{equation}
T_{RS}=T_b + \alpha_1 R^2  + \alpha_2 R^3 (k/M^3) + \ldots , 
\end{equation}
where the $R$'s correspond to curvature quantities, the $\alpha_i$
correspond to coefficients of the derivative expansion, and $M$ is the
five-dimensional Planck scale.\footnote{An investigation of domain
wall solutions in the presence of higher derivative curvature terms
can be found in Ref.~\cite{corradini}.}  Therefore, even if
$T_{RS}=O(\mbox{TeV})$, one may in principle have a $T_b$ that is much
larger, depending upon the precise nature of $\alpha_i$ and $R$.  In
that case, even without the ``confining'' effect of the orbifold fixed
point, the brane would not ``fatten'' at a TeV scale.  On the other
hand, the brane may fatten such that there may be associated collider
signatures and an amelioration of the hierarchy problem.

In this work, we assume that the RS brane can be treated as an
idealized thin domain wall and neglect any possibility of brane
fattening.  As a partial follow-up to the work of Ref.\cite{ADDK}, we
investigate both the gravitational and collider experimental prospects
for this scenario. We find that although there is no collider
phenomenology, the corrections to Newton's law distinguish the
$\overline{\rm RS}$ scenario from the ADD scenario.  We find that not
only the functional behavior of the gravitational correction is
different but the magnitudes are sufficiently different to distinguish
the two scenarios.  Indeed, identifying the scale $R$ of the extra
dimension in ADD with the AdS horizon length $1/k$ of $\overline{\rm
RS}$, we show that the $\overline{\rm RS}$ gravitational correction is
much larger at the reach of the upcoming experiments. Owing to the
larger number of moduli fields present in the the ADD scenario than in
the $\overline{\rm RS}$ scenario, other signatures may distinguish the
two scenarios.  However, these more model dependent questions will not
be addressed in this paper.

We first calculate the Newtonian potential generated by a point source
of mass \( m_{source} \) localized on the Planck brane.  Explicitly,
we consider the effects of the KK tower on the effective
four-dimensional gravitational potential following the procedure of
Garriga and Tanaka \cite{garrigatanaka}.  The details of this
calculation can be found in Ref.~\cite{garrigatanaka}, and several of
the relevant results (including a derivation of the effective action
to fix our notation and conventions) are listed in the Appendix. To
determine for completeness the correction associated with an infrared
cutoff in the AdS space, we assume the dimension transverse to the
Planck brane is an $S^1/Z_2$ orbifold, and place a cutoff brane at the
other orbifold fixed point.  We then examine the limiting behavior as
the distance $L$ between the orbifold fixed points is taken to
infinity.  In practice, it would of course be extremely difficult to
stabilize a cutoff brane at macroscopic
distances.\footnote{Furthermore, arranging a stabilization mechanism
for the radion would also obviate one of the original motivations for
having the warp factor ``truncate'' the volume of the extra dimension
because in some sense the brane tension in the RS model is meant to
replace the role of the bulk stabilizing potential.}  The potential
can be written explicitly as a sum over terms involving Bessel
functions:
\begin{eqnarray}
V=-G_N\frac{m_{source}}{r}\left[1+\frac{4}{3}\sum_{n=1}e^{-ka_nr}
\left[\left(\frac{Y_1(a_n)}{Y_1(a_ne^{kL})}\right)^2-1\right]^{-1}\right],
\end{eqnarray}
in which $a_n=m_n/k \approx n\pi e^{-kL}$. However, for qualitative
understanding it is important to note that the light KK modes (with \(
m_n \lsim k \)) couple to the matter on the brane with a strength
\(g_c \sim \sqrt{m_n/k} \) relative to the zero mode, while the heavy
KK modes (with \(m_n \grtsim k \)) couple with the same strength as
the zero mode.  The light states thus are responsible for the $1+1/(k
r)^2$ multiplicative correction to the Newtonian potential, while the
heavy states provide a multiplicative correction of the form
$1+1/(kr)$. Hence there is a transition between the two regions, as
first noted by Ref. \cite{ADDK}.  To obtain a simple analytic
expression for the potential, the two contributions must be patched
together in the summation of the KK modes. We first interpolate
between the two regions using a step-function approximation at the
cutoff \(q_c \sim O(k) \); we refer the reader to the Appendix for the
details.  We obtain
\begin{eqnarray}
V & = & -G_N\frac{m_{source}}{r} \left\{
 1+\frac{2}{3k^{2}r^{2}}-e^{-q_{c}r} \left[ \frac{2}{3k^{2}r^{2}}+\frac{2}{3(\frac{k}{q_{c}})kr}-\frac{4}{3\pi
 kr}+\left(\frac{2}{3}-\frac{\pi }{3(\frac{k}{q_{c}})}\right)e^{-kL} \right]
 \right. \nonumber \\ & - &
 \left. \frac{1}{kr}\frac{4}{3\pi }e^{-Mr}+\ldots
 \right\}\label{eq:bestapprox}
\end{eqnarray}
where $G_N$ is the four-dimensional Newton's constant, and we have kept
only the leading terms in \( e^{-kL} \) suppression. Note that terms
that depend explicitly on the ``light
mode-cutoff'' \( q_{c} \) in Eq. (\ref{eq:bestapprox}) are sensitive to 
our step function treatment of the gravitational coupling between the
states with \( m_{n} \lsim q_{c} \) and \( m_{n} \grtsim q_{c} \). Note
in particular that since \( e^{-q_{c}r} \) is sensitive to the value of
\( q_{c} \), we expect the exact form of this exponential to be an
artifact of the abrupt change in the coupling \( g_{c} \) of the KK
states between \( m_{n} \lsim q_{c} \) and \( m_{n} \grtsim q_{c} \) to
the matter on the brane. An argument for there being no exponential
suppression is presented in Ref.~\cite{german}.

Note that even in the \( L\rightarrow \infty  \) limit, the \(
e^{-q_cr}/kr \) term survives. Presumably, in the original
calculation of RS, this term was neglected due to the fact that \( k \)
was near the Planck scale. Note that
in the limit that \( kr\rightarrow 0 \) with $q_c=O(1)k$,  
\begin{eqnarray}
\label{eq:potlimit}
V & = & -G_N\frac{m_{source}}{r} \left[ 1+\frac{4}{3\pi }\frac{1}{kr}(1-e^{-Mr})+\frac{1}{3(\frac{k}{q_{c}})^{2}}-\frac{4}{3(\frac{k}{q_{c}})\pi }+O(kr)\right]
\end{eqnarray}
showing that the theory becomes five dimensional.\footnote{Note that
Eq. (\ref{eq:potlimit}) agrees with Eq. (3.38) of \cite{generalstudy} in
the $kr\ll 1$ limit.} Hence, as pointed out
by Ref.~\cite{ADDK}, the behavior for the Newtonian potential makes a
smooth transition between two and one extra flat dimensional
behavior. As we shall see, this is consistent with the consideration
in the momentum space applicable for colliders.

%%%%%%%%%%%%%%%%%%%%%%%%%%%%%%%%%%%
%%%%%%%%%%%%%%%%%%%%%%%%%%%%%%%%%%%
\begin{figure}[t]
\centering \leavevmode\epsfxsize=350pt \epsfbox{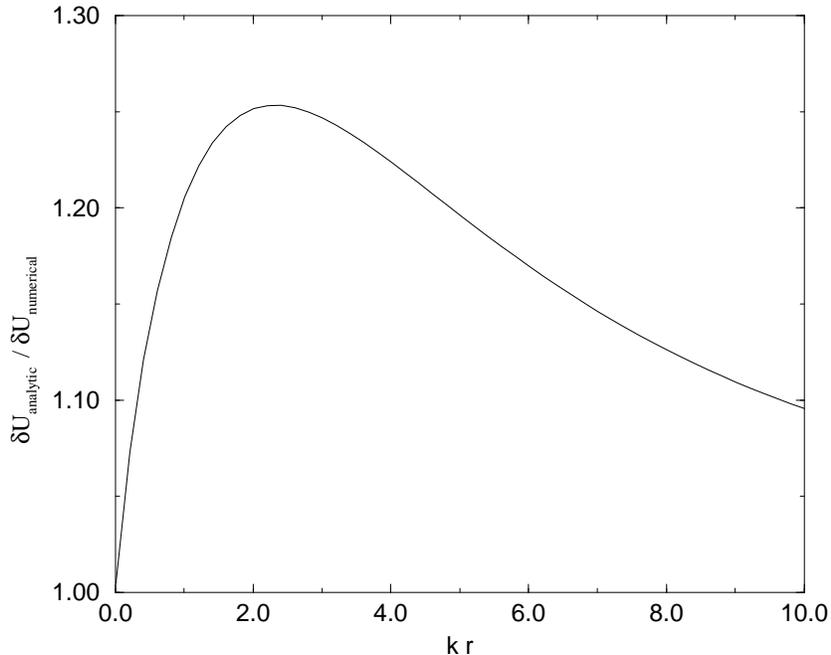}
\caption[fig1]{ A comparison of (\ref{eq:hooman}) with
(\ref{eq:bestapprox}) is given.  The Newtonian potential is given by
$-(G_N m_{source}/r)(1+\delta U_X)$ where $\delta U_{analytic}$
is given by Eq. (\ref{eq:bestapprox}) and $\delta U_{numerical}$ is
given by Eq. (\ref{eq:hooman}).}
\end{figure}
%%%%%%%%%%%%%%%%%%%%%%%%%%%%%%%%%%%
%%%%%%%%%%%%%%%%%%%%%%%%%%%%%%%%%%%

One can obtain a perhaps more accurate approximate expression for the
potential using the approximation
\begin{equation}
Y_{1}^{2}(a_{n}e^{kL})\approx \frac{2e^{-kL}}{\pi
a_{n}}\frac{Y_{1}^{2}(a_{n})}{J_{1}^{2}(a_{n})+Y_{1}^{2}(a_{n})}.
\label{eq:besselapprox}
\end{equation}
This formula can be obtained by approximating \(
J_{1}(a_{n}e^{kL}) \) and \(Y_{1}(a_{n}e^{kL}) \) as the leading
sinusoidal function. After taking the $kL\rightarrow \infty$ limit,
this results in
\begin{equation}
\label{eq:hooman}
V=-G_N\frac{m_{source}}{r} \left[ 1+\frac{8}{3\pi
^{2}}\int ^{\infty
}_{0}\frac{da}{a}\frac{e^{-akr}}{J^{2}_{1}(a)+Y^{2}_{1}(a)} \right],
\end{equation}
which is particularly useful for evaluating the Newtonian
potential corrections numerically.   Note that unlike
\eqr{eq:bestapprox}, there is no artificial separation between the
light ($m_n \lsim k$) and heavy ($m_n \grtsim k$) KK states.  
\eqr{eq:bestapprox} matches \eqr{eq:hooman} most closely
when \( q_{c}=2k/\pi  \). The ratio of the numerical values of the
correction $\delta U$ (where $\delta U$ is given by
$V=(G_Nm_{source}/r)[1+\delta U]$) of the two equations are shown in Fig.1.

Let us now compare this with the predictions of the ADD model for two
extra dimensions. Systematic corrections to the Newtonian potential have
been worked out explicitly in \cite{sfetsos,greek}; we use the results of
Ref. \cite{greek} and list the relevant expressions in Table \ref{tab1}
under the heading ``ADD.'' In the table, we take \( q_{c}=2k/\pi \) since
that gives the closest match to the numerical approximation of the
correction.  We have assumed that the two extra dimensions of the ADD
scenario each has a length of \( 2\pi R \). Note that the small \( r/R \)
expansion can be obtained through the method of images.  It is interesting
to note that there is no $R/r$ correction in this potential, even though
an integration approximation of the summation would suggest that there be
one.

%%%%%%%%%%%%%%%%%%%%%%%%%%%%%%%%%%%
%%%%%%%%%%%%%%%%%%%%%%%%%%%%%%%%%%%
\begin{figure}[t]
\centering \leavevmode\epsfxsize=350pt \epsfbox{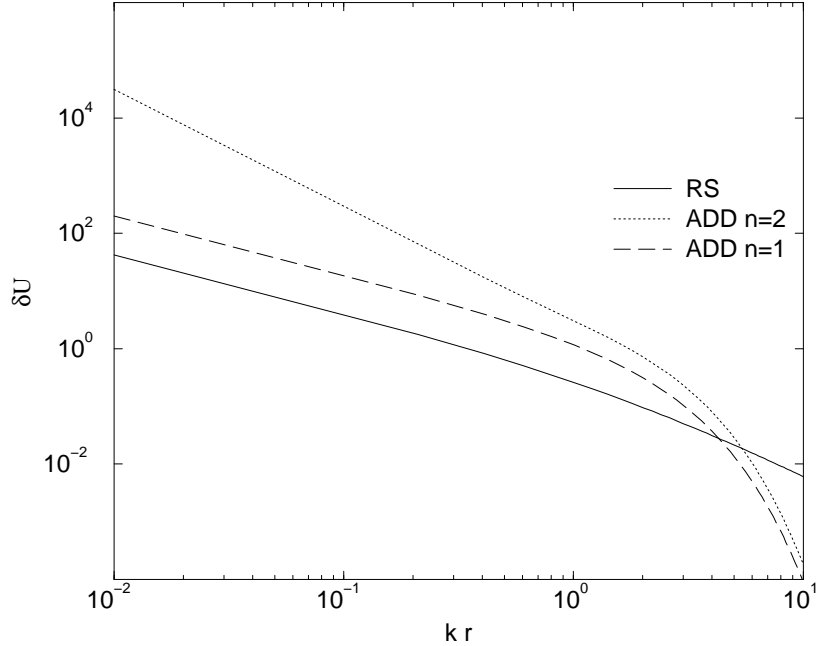}
\caption[fig1]{\label{mainresult} The corrections to the Newtonian
potential are plotted as a function of a scaled radius for the ADD
scenario with one and two extra dimensions and the $\overline{\rm RS}$
scenario. The radius $R$ of the ADD scenario is identified with the
AdS horizon $1/k$.}
\end{figure}
%%%%%%%%%%%%%%%%%%%%%%%%%%%%%%%%%%%
%%%%%%%%%%%%%%%%%%%%%%%%%%%%%%%%%%%

In Fig.~\ref{mainresult} we see the comparison between the
$\overline{\rm RS}$ and the ADD scenarios for $R=1/k$.  In addition to
the two-extra-dimension scenario which is well motivated from the point
of view of the hierarchy problem, we will also consider the one extra
dimension ADD scenario with the compactification scale $R$ set to $1/k$,
corresponding to a five dimensional Planck scale of $10^8$
GeV. Specifically, we plot $\delta U$ for the ADD model with two extra
dimensions ($n=2$) and with one extra dimension ($n=1$).  For the two extra
dimensions case, we use
\begin{equation} 
\delta U  =  2\sum
_{n_{1},n_{2}}e^{-\sqrt{n_{1}^{2}+n_{2}^{2}}\frac{r}{R}}
\end{equation}
while for the one extra dimension case
\begin{equation}
\delta U= \frac{2}{e^{r/R} -1}
\end{equation}
The figure compares these to the $\delta U$ for the $\overline{\rm
RS}$ case given by \eqr{eq:hooman}.  It shows that for $k \sim R^{-1}
\sim 10\, ({\rm mm})^{-1}$ and $r=1$ mm, multiplicative corrections of
$1+O(10^{-2})$ arise from the $\overline{\rm RS}$ scenario which may
be measured in the upcoming set of experiments (for a recent review,
see Ref.~\cite{Long:1999dk}) while the corrections arising from the ADD
scenarios are too small to be detected.  Therefore, the ADD and
$\overline{\rm RS}$ scenarios can be clearly distinguished both at
short and long distances (compared to the ``compactification scale''
$1/k$) through not only the functional behavior but also the
amplitudes.

%%%%%%%%%%%%%%%%%%%%%%%%%%%%%%%%
%%%%%%%%%%%%%%%%%%%%%%%%%%%%%%%
Let us now consider the collider signatures of the $\overline{\rm
 RS}$ model. For a typical
graviton mediated \( s \)-channel scattering process in $\overline{\rm
 RS}$ at the center
of mass energy \( \sqrt{s}\gg k \), we have for the scattering
amplitude \( A \) 
\begin{equation}
A\sim s^{2}\sum _{n}\frac{ig_{c}^{2}(n)}{s-m_{n}^{2}}\end{equation}
where we have explicitly denoted the \( n \) dependence of \( g_{c}
\). As in the gravitational potential section, we take the UV cutoff
to be at the 5-D Planck scale \( M \), at which we expect a more
fundamental theory to take over. Since we are making order of
magnitude estimates, we can approximate the summation as an integral,
and divide the KK states into the light (\( m_{n}\lsim k \)) and heavy
(\( m_{n} \grtsim k \)) groups as done in the calculation of the Newtonian
potential.  We find
\begin{equation}
A\sim \frac{s}{M^{2}_{pl}}\left[
\frac{8 \pi \sqrt{s}}{k}+i \frac{16s}{kM}\right],
\end{equation}  
 where we have used the approximations \( k\ll \sqrt{s} \), \( kL\gg 1 \),
\( \sqrt{s}\ll M \). Hence the values \( s\sim 10^{6}\textrm{GeV}^{2} \) and
\( k\sim 10^{-12} \) GeV relevant for collider experiments imply \( A\sim
(10^{-14}+i10^{-19}) \) where the real part corresponds to the resonance
channel. The easiest way to see that this number is too small for
observable collider phenomenology is to compare this to the amplitude
obtained in the ADD scenario with two extra dimensions.
With the cutoff scale at \( \Lambda =10 \) TeV in the ADD scenario, the
amplitude is
\begin{equation}
A_{ADD}\sim \frac{-is^{2}}{\Lambda ^{4}}\ln \left(\frac{\Lambda
^{2}}{s}\right)\sim -i10^{-4}.
\end{equation}
Hence, even with the resonance enhancement of \( A \) taken into account,
there is no observable phenomenology for the RS scenario even with \( k \)
as small as \( 10^{-3} \) eV.\footnote{Similarly, there will be no collider
signatures for the one extra dimension ADD model with $R\sim 0.1\,{\rm mm}$.}

In summary, we considered the uncompactified RS model of localized
gravity in the limit where the warp scale $k$ is small, the $\overline{{\rm
RS}}$ model, and studied its possible experimental consequences.  We
showed that for $k \sim 10^{-3}$ eV, this model could have measurable
contributions to deviations from Newtonian gravity, at near future
levels of experimental sensitivity.  Naturalness arguments may favor $k$
to be of order the fundamental scale.  However, a realistic fundamental
theory of quantum gravity is as yet unknown, and the only candidate,
string theory, does not yield any single RS picture.  Hence, without any
rigorous requirements at hand, we treat $k$ as a model parameter.
Nonetheless, we point out that this low value of $k$ yields an effective
energy scale of $O$(TeV) for the four-dimensional boundary
corresponding to our visible universe, a scale which may be useful for
resolving the hierarchy problem.  This leaves open the possibility that
perhaps in some more fundamental picture, low energy supersymmetry
breaking or other weak scale physics may require an effective $k \sim
10^{-3}$ eV.

We studied the contribution of the graviton KK modes to the
four-dimensional gravitational potential of a source mass, resulting
in a deviation $\delta U$ from Newtonian gravity. We not only verified
the $kr \gg 1$ and $kr \ll 1$ behavior first argued by
Ref. \cite{ADDK}, but also obtained expressions for the intermediate
regime.  We found that for $k r \grtsim 1$, the $\overline{{\rm RS}}$
model results in values of $\delta U$ that are about two orders of
magnitude larger than the corresponding ADD values for one and two
extra dimensions of size $R \sim k^{-1}$.  This makes the
$\overline{\rm RS}$ model distinguishable from the ADD scenarios in
the near future gravitational experiments. We finally noted that the
$\overline{{\rm RS}}$ model does not yield collider signatures for
$\sqrt{s} \gg k \sim (0.1\,\rm mm)^{-1}$.  In this regime, the theory
is basically that of ADD with one extra dimension of tenth of a
millimeter size and with a fundamental scale $M \sim 10^8$ GeV, which
is far beyond the reach of present and foreseeable colliders.

\acknowledgments{ It is a pleasure to thank J. Liu for conversations
regarding his papers and M. Peskin for discussions regarding the
corrections to Newton's Law.  We gratefully acknowledge conversations
with K. Choi, S. Kachru, J. Lykken, S. Rigolin, T. Rizzo, M. Schmaltz,
R. Sundrum, D. Waldram, and J. Wang.  Finally, we thank the Aspen Center for
Physics, where this work was originated, for hospitality. The work of
D.J.H.C. and L.E. was supported in part by the Department of Energy.
The work of H.D. is supported by the Department of Energy, contract
DE--AC03--76SF00515.  }
%%%%%%%%%%%%%%%%%%%%%%%%%%%%%%%%%%%% 

%\pagebreak
%%%%%%%%%%%%%%%%%%%%%%%%%%%%%%%%%%%%%%%%%
\begin{table}[t]
\vspace{0.3cm}
{\centering \begin{tabular}{|c|p{4.7in}|}
\hline
& \\ 
{           }&
{  Corrections To Newton's Law ~~~\( V\equiv
-G_{N}\frac{m_{source}}{r}(1+\delta U) \)}\\
&
\\
\hline 
\hline 
{ $\overline{\rm RS}$ }&
{   \protect{ \( \begin{array}{lll}
\\
&\delta U=\frac{2}{3k^{2}r^{2}}(1-e^{-\frac{2k}{\pi
}r})-\frac{1}{kr}(\frac{4}{3\pi }e^{-Mr})+O(e^{-2kL})& \\
\\
\\

kr\gg 1: &  \delta U=\frac{2}{3k^{2}r^{2}}+O(e^{-kr}) &  \\
\\
kr\ll 1: & \delta U=\frac{4}{3\pi
}\frac{1}{kr}(1-e^{-Mr})+O(1) & 
\\ & &
\end{array} \)}}\\
\hline 
{  ADD}&
{  \( \begin{array}{lcll}
\\
&\delta U & = &2 \sum_{n_{1},n_{2}}e^{-\sqrt{n_{1}^{2}+n_{2}^{2}}\frac{r}{R}} \\ 
\\
\\
\frac{r}{R}\gg 1: & \delta
U &= & 4e^{-\frac{r}{R}}+O(e^{-\sqrt{2}\frac{r}{R}}) \\
\\
\frac{r}{R}\ll 1: & \delta U & =
&-1+\pi(\frac{R}{r})^{2}+\frac{1}{2\pi^2}\frac{r}{R} \sum _{k=0}^{\infty }\sum
_{l=1}^{\infty }[(\frac{r}{2\pi R})^{2}+k^{2}+l^{2}]^{-3/2}
\\
\\
& & \approx & -1+\pi(\frac{R}{r})^{2}+\frac{2(2.24)}{(2\pi
)^{2}}\frac{r}{R}+O(\frac{r^{2}}{R^{2}}) 
\\ & & &
\end{array} \)}\\
\hline 
\end{tabular}  \par}
\vspace{0.3cm}
\caption[tab1]{\label{tab1}A comparison of the analytic expressions for the
corrections to the Newton's Law for the small warping Randall-Sundrum
model and the ADD scenario with two extra dimensions.} 
\end{table}
%%%%%%%%%%%%%%%%%%%%%%%%%%%%%%%%%%%%%%%%%%%%%%%%%%%%%%%%%%%%%

\bibliographystyle{prsty}

%%%%%%%%%%%%%%%%%%%%%%%%%%%%%%%%%%
\appendix
\section{Effective Action}
%%%%%%%%%%%%%%%%%%%%%%%%%%%%%%%%%%
To fix our notation, let us review the construction of the
four-dimensional effective action presented in Ref.~\cite{RSII}.
Consider the perturbed metric
\begin{equation}
ds^{2}=f(z)\left[\eta_{\mu \nu }
+h_{\mu \nu }(x)\right]dx^{\mu }dx^{\nu }-dz^{2},
\end{equation}
where the transverse dimension is parameterized by \( z \) (the Planck
brane is located at $z=0$), and the warp factor is 
\begin{equation}
f(z)=e^{-2k|z|}.
\end{equation}
We then find the effective Lagrangian for the metric perturbation to be
\begin{equation}
\Delta [\sqrt{g}(R+2\Lambda )]=-\frac{f}{2}\left[\frac{-1}{2}h^{\mu \nu
}h_{\mu \nu \, \, \, ,\alpha }^{\, \, \, ,\alpha }+h^{\mu \nu
}h^{\lambda }_{\, \, \nu ,\lambda \mu }-h^{\mu \nu }h_{,\mu \nu
}+\frac{1}{2}hh_{,\mu }^{\, \, \, ,\mu }\right] 
+ \frac{f^2}{4}\left[ h^{\mu \nu}_{\,
\, \, ,4} h_{\mu \nu ,4} - h_{,4}^2 \right]
\end{equation}
where $R$ is the five-dimensional Ricci scalar, $\Lambda$ is the bulk
cosmological constant, and the indices (which are contracted with the
Minkowski metric $\eta_{\mu\nu}$) run over $0,1,2,3$, the
coordinates parallel to the Planck brane.  In the transverse,
traceless gauge, we simply have
\begin{equation}
\Delta [\sqrt{g}(R+2\Lambda )]=\frac{-f}{4}h^{\mu \nu }h_{\mu \nu \, \, \, \, \, ,\alpha }^{\, \, \, \, ,\alpha }\end{equation}
Hence, the graviton action $S_G$ coupled minimally to the matter action \( S_{M} \)
is given by
\begin{equation}
S=S_{G}+S_{M},\end{equation}
where
\begin{equation}
S_{G}=\frac{M^{3}}{4}\int d^{5}xfh^{\mu \nu }h_{\mu \nu \, \, \, \, \, ,
\alpha }^{\, \, \, \, ,\alpha }.\end{equation}
 Let us decompose the graviton in terms of 4-dimensional modes:
\begin{equation}
h_{\mu \nu }=\xi \sqrt{\frac{\pi }{L}} \sum_n H^{(n)}_{\mu \nu }\chi ^{(n)},\end{equation}
 in which 
\begin{equation}
\partial _{z}(f^{2}\partial _{z}\chi ^{(n)})+m_{n}^{2}f(z)\chi ^{(n)}=0\end{equation}
\begin{equation}
\partial _{z}\chi ^{(n)}(0)=\partial _{z}\chi ^{(n)}(L)=0\end{equation}
with the normalization
\begin{equation}
\int ^{L}_{-L}dzf(z)\chi ^{(n)}(z)\chi ^{(m)}(z)=\frac{L}{\pi
}\delta_{nm}. \end{equation} 
Note that $\xi$ is an arbitrary normalization parameter.  Hence, we
find for the four dimensional effective action of gravity 
\begin{equation}
S_{G}=\frac{\xi^2 M^{3}}{4}\int d^{4}x\{H^{\mu \nu }_{(n)}
H_{\mu \nu ,\alpha }^{(n)\, \, \, \, ,\alpha }
+m_{n}^{2}H^{\mu \nu }_{(n)}H_{\mu \nu }^{(n)}\}.\end{equation}
We can also expand the matter action to give
\begin{eqnarray}
S_{M} & = & S_{M}^{(0)}+\int d^{4}x\frac{\delta S}{\delta g_{\mu \nu }(x)}h_{\mu \nu }(x,z=0)+\ldots\\
 & = & S_{M}^{(0)}-\sum_n \int d^{4}x\frac{\sqrt{g(z=0)}}{2}T^{\mu \nu }_{brane}H_{\mu \nu }^{(n)}\chi ^{(n)}(z=0)\xi \sqrt{\frac{\pi }{L}}+\ldots
\end{eqnarray}
where the $\ldots$ include comparable coupling terms induced from the
fact that the brane is bent in the presence of matter in the
transverse traceless gauge.
The effective action is given by 
\begin{equation}
\delta S=\frac{\xi^2 M^3}{4} \int d^{4}x\{H^{\mu \nu }_{(n)}H_{\mu \nu
,\alpha }^{(n)\, \, 
\, \, ,\alpha }+m_{n}^{2}H^{\mu \nu }_{(n)}H_{\mu \nu
}^{(n)}\}-\frac{1}{2}\int d^{4}xT^{\mu \nu }_{brane}H_{\mu \nu
}^{(n)}\chi ^{(n)}(z=0)\xi \sqrt{\frac{\pi }{L}} + \ldots.
\end{equation}
To see the Planck scale suppression of the couplings, it is convenient
to normalize the kinetic term for the gravitons such that $H_{\mu
\nu}$ has dimension 1 and the propagator resembles a scalar
propagator.  Hence, we let
\begin{equation}
\xi^2=\frac{4}{M^3}
\end{equation}
Now, we have for \( z>0 \),
\begin{equation}
\chi ^{(n)}(z)=\frac{1}{N_{n}}[J_{2}(a_{n}e^{kz})+\alpha _{n}Y_{2}(a_{n}e^{kz})]e^{2kz}\end{equation}
 with 
\begin{equation}
\alpha _{n}=\frac{-J_{1}(a_{n})}{Y_{1}(a_{n})}\end{equation}
where 
\begin{equation}
a_{n}=\frac{m_{n}}{k},\end{equation}
 and the mass condition is given by 
\begin{equation}
\frac{J_{1}(a_{n}e^{kL})}{J_{1}(a_{n})}=\frac{Y_{1}(a_{n}e^{kL})}{Y_{1}(a_{n})}.\end{equation}
To get the normalization constant \( N_{n} \), we use
\begin{equation}
\frac{2}{N^{2}_{n}}\int ^{L}_{0}dz [ J_{2}(a_{n}e^{kz})+\alpha
 _{n}Y_{2}(a_{n}e^{kz})]^{2}e^{2kz}=\frac{L}{\pi}.\end{equation}
 Writing 
\begin{equation}
Z_{n}(x)\equiv J_{n}(x)+\alpha Y_{n}(x),\end{equation} for any constant
$\alpha$ and real argument $x$, we can use the Bessel function identity
\begin{equation}
\int dxxZ^{2}_{n}(ax)=\frac{x^{2}}{2}[Z^{2}_{n}(ax)-Z_{n+1}(ax)Z_{n-1}(ax)]\end{equation}
for any integer $n$ and the mass shell condition to get 
\begin{equation}
N^{2}_{n}=\frac{\pi}{kL}\left[e^{2kL}Z^{2}_{2}(a_{n}e^{kL})-Z^{2}_{2}(a_{n})\right].\end{equation}
 Therefore, the factor \( \chi ^{(n)} \) which determines the matter
 coupling to gravity is given by 
\begin{eqnarray}
\chi ^{(n)}(0) & = & \sqrt{\frac{kL}{\pi }}\frac{1}{\sqrt{\left[
\frac{Y_{1}(a_{n})}{Y_{1}(a_{n}e^{kL})}\right] ^{2}-1}}
\label{eq:wavefn}
\end{eqnarray}
 in which we used the Bessel function identity 
\begin{equation}
J_{\nu }(x)Y_{\nu +1}(x)-J_{\nu +1}(x)Y_{\nu }(x)=-\frac{2}{\pi x},\end{equation}
 and the definition of \( \alpha _{n} \). The coupling of the gravitational
modes to matter is then given by 
\begin{equation}
\delta S_{M} = \int d^{4}xT^{\mu \nu }g_{c}^{(n)}H_{\mu \nu }^{(n)},
\end{equation}
 where we have defined the coupling 
\begin{equation}
g^{(n)}_{c}\equiv \frac{-\xi}{2} \sqrt{\frac{\pi}{L}}
\chi^{(n)}(z=0)=\frac{-1}{M^{3/2}}\sqrt{\frac{\pi}{L}} \chi^{(n)}(z=0).
\end{equation}
%%%%%%%%%%%%%%%%%%%%%%%%%%%%%%%%%%%
\section{Newtonian Potential}
%%%%%%%%%%%%%%%%%%%%%%%%%%%%%%%%%%%%
We now calculate the Newtonian potential generated by a point source of
mass \( m_{source} \) localized on the brane, following the procedure of
Garriga and Tanaka,\cite{garrigatanaka}. We find 
\begin{eqnarray}
V & = & -\frac{1}{2M^3} m_{source}\left\{\frac{2}{3}\sum
_{n}\frac{e^{-m_{n}r}}{4\pi r}\frac{\pi \chi
^{2}_{n}(0)}{L}-\frac{k}{24\pi r} \right\}
\label{eq:potential}
\end{eqnarray}
where the factor of \( 2/3 \) can be attributed to the tensorial
character of gravity and brane bending. Note that since the bound
state mode \( \chi _{0} \) is normalized as
\begin{equation}
\chi _{0}=\sqrt{\frac{kL}{\pi (1-e^{-2kL})}},
\end{equation} 
the recovery of Newton's law for this mode requires
\begin{equation}
\frac{M_{pl}^2}{8 \pi} = \frac{2 M^3}{k}
\end{equation}
to first order in $e^{-k L}$ (where $M_{pl}^2\equiv 1/G_N$).  Hence, we have
\begin{equation}
g_c = -\frac{\sqrt{2}}{(M_{pl}/\sqrt{8 \pi})} 
\frac{1}{\sqrt{ \left[\frac{Y_1(a_n)}{Y_1(a_n e^{kL})} \right]^2 -1 }} 
\label{eq:defcoupling}
\end{equation}

 Let us examine \( g_{c} \) in various limits, 
\begin{enumerate}
\item \( a_{n}\ll 1 \): In this case\footnote{The case of \(
a_{n}e^{kL}\ll 1 \) corresponds to just the zero mode solution.}
\begin{equation}
\label{eq:lightm}
g_{c}=-\frac{2\pi
\sqrt{2}}{M_{pl}}\sqrt{\frac{m_{n}}{k}}e^{-\frac{kL}{2}}+O(e^{-3kL})
\end{equation}

\item \( a_{n}\gg 1 \): In this case
\begin{equation}
\label{eq:heavym}
g_{c}=-\frac{4 \sqrt{\pi}}{M_{pl}} \left\{ e^{-\frac{kL}{2}}
\left[ 1-\frac{15}{128} \left(\frac{k}{m_{n}}\right)^{2}
\right]+e^{-\frac{3kL}{2}}
\left[ \frac{1}{2}-\frac{45}{256} \left(\frac{k}{m_{n}}\right)^{2} \right]
+
O\left(e^{-\frac{5kL}{2}} \right) \right\}
\end{equation}   

\end{enumerate}
where we have kept the leading term correction in \( e^{-kL}<1 \)
suppression. For both cases it is valid to approximate \( a_{n}\approx
[n+1/4 - \epsilon(n) ]\pi e^{-kL} \) where \( \epsilon(n) \approx 3/[8
\pi^2 (n+1/4)] \). In the expressions above, and from now on, we
neglect all dependence on $\epsilon(n)$.\footnote{Since the
$\epsilon(n)$ dependence can come into the gravitational potential
expression through multiplicative corrections of $1+O(\epsilon)$
through $\chi^{(n)}$ and $m_n$, we see that the KK tower summation
will be corrected at most by a multiplicative factor of
$1+O(\epsilon(1))$.}  Note that the $(k/m_n)^2$ terms in \eqr{eq:heavym}
can also be neglected.

%For
%the second case, we have also used the fact that the mass condition 
%\( J_{1}(a_{n}e^{kL})/J_{1}(a_{n})=Y_{1}(a_{n}e^{kL})/Y_{1}(a_{n}) \)
%implies \( a_{n}e^{kL}-a_{n} \approx j\pi \) for some large integer \(
%j \) .

In summing over the KK states, we split the states into two groups,
 one with \( m_{n}\leq q_{c} \) (\( q_{c} \) is a scale with order of
 magnitude \( O(q_{c})=k \)) and coupling \eqr{eq:lightm}, and the
 other with \( m_{n}>q_{c} \) and coupling \eqr{eq:heavym}. In
 reality, there is some smooth function approximated by
 \eqr{eq:defcoupling} that should be used instead, but for the
 purposes of carrying out the summation explicitly, we have used
 \eqr{eq:lightm} for modes \( n\leq -1/4+q_c e^{kL}/k\pi \) and
 \eqr{eq:heavym} for the rest (essentially a step function
 treatment). We have also taken the cutoff for the \( m_{n}> q_c \)
 modes to be \( M \) (i.e. restricted to modes \( n<-1/4+M e^{kL}/k\pi
 \) ). Carrying out the summation (without using an integral
 approximation), we obtain \eqr{eq:bestapprox}
\begin{eqnarray}
V & = & -\frac{1}{M^{2}_{pl}}\frac{m_{source}}{r} \left\{
 1+\frac{2}{3k^{2}r^{2}}-e^{-q_{c}r} \left[ \frac{2}{3k^{2}r^{2}}+\frac{2}{3(\frac{k}{q_{c}})kr}-\frac{4}{3\pi
 kr}+(\frac{2}{3}-\frac{\pi }{3(\frac{k}{q_{c}})})e^{-kL} \right]
 \right. \nonumber \\ & - &
 \left. \frac{1}{kr}(\frac{4}{3\pi }e^{-Mr})+\ldots
 \right\},
\label{eq:bestapproxapp}
\end{eqnarray}
where we have kept terms to only leading order in $O(e^{-kL})$ and
$O(e^{-Mr})$.
\end{document}